\documentstyle[aps,prl,twocolumn,floats,psfig,epsfig]{revtex}

\newcommand{\fd}{fluc\-tu\-a\-tion-dis\-si\-pa\-tion }
\newcommand{\beq}{\begin{equation}}
\newcommand{\beqa}{\begin{eqnarray}}
\newcommand{\eeq}{\end{equation}}
\newcommand{\eeqa}{\end{eqnarray}}

\newcommand{\dpar}{\partial}
\renewcommand{\d}{{\rm d}}
\def\gt{{\tilde{g}}}
\def\vt{{\tilde{v}}}

\begin{document}
\draft
\twocolumn[\hsize\textwidth\columnwidth\hsize\csname@twocolumnfalse\endcsname
\title{Aging and fluctuation-dissipation ratio for the diluted Ising Model}
\author{Pasquale Calabrese and Andrea Gambassi}
\address{{Scuola Normale Superiore and INFN,
Piazza dei Cavalieri 7, I-56126 Pisa, Italy. 
}
\\
{\bf e-mail: \rm {\tt calabres@df.unipi.it}, {\tt andrea.gambassi@sns.it} }}
\date{\today}

\maketitle

\begin{abstract}
We consider the out-of-equilibrium, purely relaxational dynamics of a weakly 
diluted Ising model in the aging regime at criticality. 
We derive at first order
in a $\sqrt{\epsilon}$ expansion the two-time response and correlation 
functions for vanishing momenta.
The long-time limit of the critical fluctuation-dissipation ratio is computed
at the same order in perturbation theory. 
\end{abstract}

]


According to universality hypothesis, critical phenomena can be described
in terms of quantities that do not depend on the microscopic details of the 
systems, but only on global properties such as symmetries,  
dimensionality etc. A question of theoretical and experimental interest is 
whether 
and how this critical behavior is altered by introducing in the systems a small
amount of uncorrelated impurities leading to models with quenched disorder.

The static critical behavior of these systems is well understood thank to the
Harris criterion~\cite{Harris-74}. It states that the addition of 
impurities to a system which undergoes a second-order 
phase transition does not change the critical behavior 
if the specific-heat critical exponent $\alpha_{\rm p}$ of the pure 
system is negative. If $\alpha_{\rm p}$ is positive, the transition
is altered. 

For the very important class of the three-dimensional $O(M)$-vector models
it is known that $\alpha_{\rm p}<0$ for $M\geq2$~\cite{PV-r}, 
and the critical behavior is unchanged in presence of weak quenched disorder.
Instead, the specific-heat exponent of the three-dimensional Ising model 
is positive~\cite{PV-r}, thus the existence of a new Random Ising Model (RIM) 
universality 
class is expected, as confirmed by Renormalization Group (RG) analyses, 
Monte Carlo simulations (MCs), 
and experimental works (see Refs.~\cite{PV-r,ff} for a comprehensive review on
the subject, and for an updated list of references).

The purely relaxational equilibrium dynamics (Model A of Ref.~\cite{HH})
of this new universality class is under intensive investigation 
~\cite{gmm-krey-77,lp-83,jos-95,fd-din,MC}. 
The dynamic critical exponent $z$
differs from the mean-field value already
in one-loop approximation~\cite{gmm-krey-77}, at variance with the pure model.
This exponent is known at three-loop level in an 
$\sqrt{\epsilon}$~\cite{jos-95} and in
fixed ($d=2,3$) dimension~\cite{fd-din} expansion, and has a value in 
good agreement with several MCs~\cite{MC}.

The out-of-equilibrium dynamics is instead less studied. The initial
slip exponent $\theta$ of the response function was determined at two-loop 
order~\cite{oj-95} and the response function only at one-loop, both for
conservative and non-conservative dynamics~\cite{kissner-92}.

In this work we focus on a different regime of 
out-of-equilibrium dynamics: the aging one. 
The relaxation of the system from an out-of-equilibrium initial state
is characterized by two different regimes:
a transient one with off-equilibrium evolution , for $t<t_R$, and a
stationary one with equilibrium evolution of fluctuations 
for $t>t_R$, where $t_R$ is the relaxation time.
In the former a dependence of the behavior of the system on initial condition
is expected, while in the latter homogeneity of time and time reversal 
symmetry~(at least in the absence of external fields) are recovered.
Consider the system in a disordered state for the initial time 
$t=0$, and quench it to a given temperature $T\geq T_c$, where $T_c$ 
is the critical temperature. Calling $\phi_{\bf x}(t)$ the order parameter
of the model, its 
response to an external field $h$ applied at a time 
$s>0$ and in ${\bf x}=0$ is given by the response function 
$R_{\bf x} (t,s)=\delta\langle \phi_{\bf x} (t)\rangle/ \delta h (s)$, where
$\langle\cdots\rangle$ stands for the mean over stochastic dynamics.
The two-time correlation function 
$C_{\bf x} (t,s)=\langle \phi_{\bf x} (t)\phi_{\bf 0} (s)\rangle$
is useful to describe the dynamics of order parameter fluctuations.

When the system does not reach the equilibrium 
(i.~e. $t_R=\infty$) all the previous functions will
depend both on $s$~(the ``age'' of the system) and $t$.
This behavior is usually referred to as aging and was
first noted in spin glass systems~\cite{review}.
To characterize the distance from equilibrium of an aging system, 
evolving at a fixed temperature $T$, the 
\fd ratio~(FDR) is usually introduced \cite{ckp-94}:
\beq
X_{\bf x}(t,s)=\frac{T\, R_{\bf x}(t,s)}{\dpar_s C_{\bf x}(t,s)} \; .
\label{dx}
\eeq
When $t,s\gg t_R$ the \fd theorem holds and 
thus $X_{\bf x}(t,s)=1$.

Recently \cite{ckp-94,lz-00,gl-000,gl-00,gl-02b,cg-02,cg-02b,hpgl-01} 
attention has been paid to the FDR, for nonequilibrium and nonglassy 
systems quenched at their critical temperature $T_c$ 
from an initial disordered state \cite{ph-02}.
The scaling form  for $R_{\bf x=0}$ was rigorously established
using conformal invariance \cite{hpgl-01,henkel-02}.
Within the field-theoretical approach to critical dynamics, 
calculations are simpler if done in momentum space, thus we
are interested in momentum-dependent response and correlation functions.
From RG arguments it is expected that they scale, for ${\bf q=0}$, as
\cite{jss-89,gl-00,cg-02,cg-02b}
\beqa
T_c R_{\bf q=0}(t,s)&=&A_R (t-s)^a (t/s)^\theta F_R(s/t) \, ,
\label{Rscalform}\\
C_{\bf q=0}(t,s)&=&A_C s (t-s)^a (t/s)^\theta F_C (s/t)\, ,\label{Cscalform} \\
\dpar_s C_{\bf q=0}(t,s)&=&A_{\dpar C}(t-s)^a (t/s)^\theta F_{\dpar C} (s/t)\,,
\label{dCscalform}
\eeqa
where $a=(2-\eta-z)/z$. 
The functions $F_C(v)$, $F_{\dpar C}(v)$ and $F_R(v)$ are universal provided
one fixes the nonuniversal normalization constant $A_R$, $A_C$, 
and $A_{\dpar C}$ to have $F_i(0)=0$.
Obviously $A_C$, $A_{\dpar C}$, $F_C(v)$, and $F_{\dpar C}(v)$ are not 
independent, in fact it holds
\beqa
A_{\dpar C}&=&A_C (1-\theta) \,,\\
F_{\dpar C}(v)&=&F_C(v)+{v\over 1-\theta}\left(F'_C(v)- {a\over1-v}F_C(v)\right)\, .
\eeqa

It has been argued that the limit 
$X_{{\bf x}=0}^\infty=\lim_{s\to\infty}\lim_{t\to\infty}X_{{\bf x}=0}(t,s)$
is a novel universal quantity of nonequilibrium critical dynamics 
~\cite{gl-000,gl-02b} and may be computed as an amplitude ratio. 
Also the analog in ${\bf q}$ space~\cite{cg-02}
\beq
{\cal X}_{{\bf q}=0}(t,s)=
{\Omega R_{{\bf q}=0}(t,s)\over \dpar_s C_{{\bf q}=0}(t,s)}=
{A_R\over A_{\dpar C}}={A_R\over A_C (1-\theta)}
\label{xinfdef}
\eeq
is universal. In Ref.~\cite{cg-02} an heuristic argument is given to show that
$X_{{\bf x}=0}^\infty={\cal X}_{{\bf q}=0}^\infty$.

In a recent work \cite{cg-02b} we evaluated $X_{{\bf x}=0}^\infty$ 
at criticality for 
$O(M)$-vector models at the second order in the $\epsilon$-expansion,
finding values in very good agreement with two- and three- dimensional
numerical simulations for the Ising model~\cite{gl-000,gl-02b}. 
We also confirmed the validity of the scaling laws~(\ref{Rscalform}) 
and~(\ref{dCscalform}) at the same order in
perturbation theory. A different model
of relaxation (Model C of Ref.~\cite{HH}) has also been studied~\cite{prep}.

The extension of this kind of investigation to disordered systems 
is very interesting because, besides giving a
check of the expected scaling laws, it predicts a new universal 
 dynamical quantity (the long time limit of the FDR) which could be measured
in MCs and could be used to identify a universality class, as in the case
of other universal quantities.

The time evolution of an $N$-component field $\varphi({\bf x},t)$ under a
purely dissipative relaxation dynamics (Model A of Ref.~\cite{HH}) 
is described by the stochastic Langevin equation
\beq
\label{lang}
\dpar_t \varphi ({\bf x},t)=-\Omega 
\frac{\delta \cal{H}_{\psi}[\varphi]}{\delta \varphi({\bf x},t)}+\xi({\bf x},t) \; ,
\eeq
where $\Omega$ is the kinetic coefficient, 
$\xi({\bf x},t)$ a zero-mean stochastic Gaussian noise with correlations
\beq
\langle \xi_i({\bf x},t) \xi_j({\bf x}',t')\rangle= 2 \Omega \, \delta({\bf x}-{\bf x}') \delta (t-t')\delta_{ij}.
\eeq
and $\cal{H}_\psi[\varphi]$ the static Landau-Ginzburg Hamiltonian \cite{PV-r}
\beq
{\cal H}_{\psi}[\varphi] = \int\!\! \d^d x \left[
\frac{1}{2} (\partial \varphi )^2 + \frac{1}{2} (r_0 +\psi({\bf x}))\varphi^2
+\frac{1}{4!} g_0 \varphi^4 \right] .
\label{lg}
\eeq
Here $\psi({\bf x})$ is a spatially uncorrelated random field with 
Gaussian distribution
\beq
P(\psi) = {1\over \sqrt{4\pi w} } \exp\left[ - {\psi^2\over 4 w}\right].
\eeq

Dynamical correlation functions, generated by Langevin equation (\ref{lang}) 
and averaged over the noise $\xi$, can be obtained by the 
field-theoretical action \cite{bjw-76} 
\beq
S_{\psi}[\varphi,\tilde{\varphi}]= \int\!\! \d t\, \d^dx 
\left[\tilde{\varphi} \frac{\partial\varphi}{\partial t}+
\Omega \tilde{\varphi} \frac{\delta \mathcal{H}_\psi[\varphi]}{\delta \varphi}-
\tilde{\varphi} \Omega \tilde{\varphi}\right].
\eeq
where $\tilde{\varphi}({\bf x},t)$ is the response field.

The effect of a macroscopic initial condition 
$\varphi_0({\bf x})=\varphi({\bf x},t=0)$ may be taken into account by 
averaging over the initial configuration
with a weight $e^{-H_0[\varphi_0]}$, where
\beq
H_0[\varphi_0]=\int\!\! \d^d x\, \frac{\tau_0}{2}(\varphi_0({\bf x})-a({\bf x}))^2.
\eeq
This specifies an initial state $a({\bf x})$ with short-range 
correlations proportional to $\tau_0^{-1}$. 

In this way all response and correlation functions may
be obtained as averages on the functional weight
$\exp\{-(S_{\psi}[\varphi,\tilde{\varphi}]+H_0[\varphi_0])\}$.
In the analysis of static critical behavior, the average over the quenched
disorder $\psi$ is usually performed by means of the replica trick~\cite{PV-r}.
If instead we are interested 
in dynamic processes it is simpler to perform directly the average
at the beginning of the calculation \cite{dedominicis-78}
\beq
\int [\d \psi] P(\psi) \exp(-S_{\psi}[\varphi,\tilde{\varphi}])=
\exp(-S [\varphi,\tilde{\varphi}])
\eeq
with the $\psi$-independent action
\beqa
S [\varphi,\tilde{\varphi}]&=&\int\!\! \d^dx \Bigg\{
\int_0^\infty\!\! \d t \, \tilde{\varphi}\left[ \dpar_t \varphi+\Omega( r_0 -\Delta)\varphi
-\Omega\tilde{\varphi}\right]\nonumber\\ 
&&
+{\Omega g_0\over 3!} \int_0^\infty\!\! \d t\, \tilde{\varphi}\varphi^3 -{\Omega^2 v_0 \over2}
\left(\int_0^\infty\!\! \d t\,\tilde{\varphi}\varphi\right)^2\Bigg\}\, ,
\label{ranaction}
\eeqa
where $v_0\propto w$.

The perturbative expansion is performed in terms of the two fourth-order
couplings $g_0$ and $v_0$ and using the 
propagators of the free theory with an initial condition at $t=0$, 
$\langle \tilde{\varphi_i}({\bf q},s) \varphi_j(-{\bf q},t) \rangle_0 = 
\delta_{ij} R^0_q(t,s)$ and 
$\langle \varphi_i({\bf q},s) \varphi_j(-{\bf q},t) \rangle_0 =\delta_{ij} C^0_q(t,s)$ 
\cite{jss-89}.
It has also been shown that $\tau_0^{-1}$ is irrelevant~(in RG 
sense) for large times behavior~\cite{jss-89}.  
From the technical point, the breaking of homogeneity in time gives rise to 
some problems in the renormalization procedure in terms of one-particle 
irreducible correlation functions~(see Ref. \cite{jss-89} and references 
therein) so all the computations are done in terms of connected functions.

\begin{figure}[t]
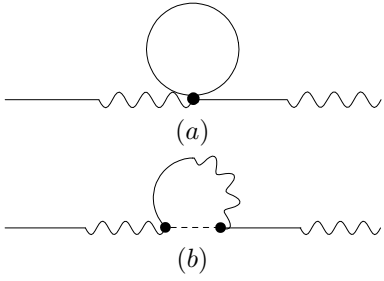

\centerline{\psfig{width=5truecm,file=random1a.eps}}
\centerline{$(a)$}
\vspace{1mm}
\centerline{\psfig{width=5truecm,file=random1b.eps}}
\centerline{$(b)$}
\vspace{2mm}
\caption{Feynman diagrams contributing to the one-loop response function.
Response functions are drawn as wavy-normal lines, whereas
correlators are normal lines. A wavy line is attached to the response field
and a normal one to the order parameter.
The dotted line is a non-local $v$-like vertex.}
\label{figR}
\end{figure}

To compute the response function at one-loop level, 
we have to evaluate the two Feynman diagrams depicted in Fig.~\ref{figR}.
In terms of them we may write
\beq
R_{\bf q}(t,s)=R^0(t,s)- {1\over 2} g_0 (a)+ v_0 (b) 
+O(g_0^2,v_0^2,g_0 v_0)\, ,
\label{Rexp}
\eeq
where we are considering the case $N=1$ (RIM universality 
class), and we
set $\Omega=1$ to lighten the notation.

In the following we report the expressions of 
Feynman diagrams at criticality ($r_0=0$ in dimensional 
regularization) for vanishing
external momentum, since we are only interested in that limit,
 and since expressions for nonzero ${\bf{q}}$ are long and not 
very illuminating.

The diagram $(a)$ in Fig.~\ref{figR} contributes also the the response 
function of 
nondisordered models, and it has been computed in Ref.~\cite{cg-02}, 
obtaining~(for $t>s$):
\beq
(a)= -N_d {1\over4} \log{t\over s} + O(\epsilon).
\eeq
where $N_d=2/[\Gamma(d/2) (4\pi)^{d/2}]$.
For diagram $(b)$ we find
\beqa
(b)&=& \int_0^\infty\!\! \d t' \, \d t'' \int\!\! {\d^d p\over(2\pi)^d}
R_0(t,t')R_{\bf p}(t',t'') R_0(t'',s)\nonumber \\
&=&{1\over (4\pi)^{d/2}}{1\over 1-d/2} {1\over 2-d/2} (t-s)^{2-d/2}\, .
\eeqa

Inserting the expression for $(a)$ and $(b)$ in Eq.~(\ref{Rexp}) and 
expanding $(b)$ at first order in $\epsilon$, one obtains
\beqa
R_B(t,s)&=&
1+\gt_0 {1\over 8} \ln {t\over s}-
{\vt_0\over2} \left[ {2\over\epsilon} +\log (t-s) +\gamma_E \right] 
\nonumber\\&&
+O(\epsilon^2,\epsilon \gt_0,\epsilon\vt_0,\gt_0^2,\vt_0^2,\gt_0 \vt_0)\,,
\eeqa
where $\gt_0=N_d g_0$ and $\vt_0= N_d v_0$.
The dimensional pole in this expression can be canceled by a multiplicative 
renormalization of both the fields $\varphi$ and $\tilde{\varphi}$.

The critical response function is then obtained setting the renormalized 
couplings at their fixed point values.
We remind that the stable fixed point of the 
RIM is of order $\sqrt{\epsilon}$ and not 
$\epsilon$~(see, e.~g.~\cite{PV-r}),
due to the degeneracy of the one-loop $\beta$ functions.
The nontrivial fixed point values at the first non-vanishing 
order (i.~e. two loops) are
\beq
\gt_0=\gt^*=4\sqrt{6\epsilon\over 53} +O(\epsilon)\,, \; 
\vt_0=\vt^*= \sqrt{6\epsilon\over 53} +O(\epsilon).
\eeq
Finally we get
\beq
R(t,s)=
1+ {1\over2} \sqrt{6\epsilon\over 53} 
\left[\ln {t\over s}-\ln (t-s) - \gamma_E\right]  
+O(\epsilon)\, ,
\eeq
that is fully compatible with the expected scaling form~(\ref{Rscalform}) with 
the well known exponents~\cite{jos-95,kissner-92}
\beq
a=  -{1\over2}\sqrt{6\epsilon\over 53}  +O(\epsilon)\quad , \quad 
\theta={1\over2} \sqrt{6\epsilon\over 53}+O(\epsilon)\;,
\label{exps}
\eeq
and the new results $F_R(x)=1 + O(\epsilon)$ and
\beq
A_R=1-{1\over2}\sqrt{6\epsilon\over 53}\gamma_E +O(\epsilon)
\; 
.
\eeq

There are five diagrams contributing to the correlation function.
Four of them are obtained by the ones of Fig.~\ref{figR}
changing one of the two external response propagators with a correlation line
(see Ref.~\cite{cg-02b} for a detailed explanation of this correspondence).
We call these four diagrams $(a_1)$, $(a_2)$, $(b_1),$ and $(b_2)$. 
The sum $(a_1)+(a_2)$ was computed in~\cite{cg-02} leading to
\beq
(a_1)+(a_2)=-{N_d\over2} s \left( \log {t\over s}+2\right)+O(\epsilon)\, .
\eeq
The sum $(b_1)+(b_2)$ is instead 
\beqa
(b_1)+(b_2)&&= {N_d \Gamma(d/2)\over(1-d/2) (2-d/2) (3-d/2)}
\times\nonumber\\&&
[ t^{3-d/2} + s^{3-d/2} - (t-s)^{3-d/2} ].
\eeqa

\begin{figure}[b]
\centerline{\psfig{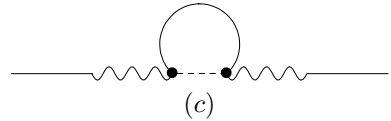}}
\centerline{$(c)$}
\vspace{2mm}
\caption{Feynman diagram contributing to the one-loop correlation function
that does not have analog in the response.}
\label{figC}
\end{figure}

The octopus diagram in Fig~\ref{figC} does not have a corresponding 
one contributing to the response function. It has the value
\beqa
(c)&=& {N_d \Gamma(d/2)\over(1-d/2) (2-d/2) (3-d/2)}
\times\\&&\nonumber
\left[{(t-s)^{3-d/2}+(t+s)^{3-d/2}\over2}-t^{3-d/2} - s^{3-d/2}\right] \, .
\eeqa
Collecting together these contributions and expanding in powers of 
$\epsilon$ we find, at 
$O(\epsilon^2,\epsilon \gt_0,\epsilon\vt_0,\gt_0^2,\vt_0^2,\gt_0 \vt_0)$,
\beqa
C_B(t,s) &=&
2s-{g_0\over2} ((a_1)+(a_2))+ v_0 ((b_1)+(b_2)+(c)) \nonumber\\
&=& 2 s+{\gt_0\over4} s \left(\log {t\over s}+2\right)
+\vt_0\left[-{2s\over\epsilon}-\gamma_E s
\right.\\&&\left.
+s+{(t-s)\log (t-s)\over2}-{(t+s) \log(t+s)\over2}\right]
.\nonumber
\eeqa
Also this expression is renormalized by canceling the dimensional pole
by means of multiplicative renormalizations of parameters and fields.
At the nontrivial fixed point for couplings we have 
\beqa 
{C(t,s)\over2}&=&
s\left\{1+{1\over2} \sqrt{6\epsilon\over 53}
\left[ \log {t\over s}+3-\gamma_E-\log (t-s)
\right.\right.\nonumber\\&&\left.\left.
+{t+s\over2s} \log {t-s\over t+s}
\right]\, \right\} + O(\epsilon) \ ,
\eeqa
which is again compatible with the expected scaling form 
(cf. Eq.~(\ref{Cscalform}))
with the same exponents of Eq.~(\ref{exps}), the non universal amplitude
\beq
{A_C\over2}=1+{1\over2} \sqrt{6\epsilon\over 53} (2-\gamma_E)+ O(\epsilon)\,,
\eeq
and the universal regular scaling function
\beq
F_C(x)=1+{1\over2}\sqrt{6\epsilon\over 53}
\left[1+{1\over2} \left(1+{1\over x}\right) \log {1-x\over1+x} \right]+ O(\epsilon)  \,.
\eeq
Note that at variance with the pure model~\cite{cg-02,cg-02b}, 
the function $F_C(x)$ receives a contribution already at one-loop order 
which should be observable in MCs.

Using Eq.~(\ref{xinfdef}) the FDR is 
\beq
{\cal X}_{{\bf q}=0}^\infty={1\over2}-{1\over4}\sqrt{6\epsilon\over 53} 
+ O(\epsilon) \; ,
\eeq
that, for $\epsilon=1$ leads to ${\cal X}_{{\bf q}=0}^\infty\sim 0.416$, 
and $\sim 0.381$ for $\epsilon = 2$. 
It would be interesting to see if this one-loop result is in
as good agreement with MCs as in the case of the pure 
model~(cf. Refs.~\cite{cg-02,cg-02b}). To this order it is not even clear 
whether randomness really changes in a sensible way
the limit of the FDR or not. Two-loop
computations and MCs could clarify this point.

For completeness we report also the FDR for finite times:
\beq
{{\cal X}_{\bf q=0}^{-1}(t,s)\over 2}=1+{1\over2}\sqrt{6\epsilon\over 53}
\left[1+{1\over2} \log {t-s\over t+s}\right] + O(\epsilon) \;.
\eeq

\end{document}